\documentstyle[aps,prl]{revtex}

\input{tcilatex}

\begin{document}
\title{Evidence for crossover from a Bose-Einstein condensate to a BCS-like
superconductor with doping in YBa$_{2}$Cu$_{3}$O$_{7-\delta }$ from
quasiparticle relaxation dynamics experiments.}
\author{J.Demsar$^{1}$, B.Podobnik$^{1}$, J.Evetts$^{2}$, G.Wagner$^{2}$ and
D.Mihailovic$^{1}$}
\address{$^{1}$Jozef Stefan Institute, 1001 Ljubljana, Slovenia\\
$^{2}$Materials Science Department, University of Cambridge, Cambridge, U.K.}
\maketitle

\begin{abstract}
Time resolved measurements of quasiparticle (QP) relaxation dynamics on the
femtosecond timescale are reported as a function of temperature and doping
in YBa$_{2}$Cu$_{3}$O$_{7-\delta }$ for ($0.48<\delta <0.1$). In the
underdoped state ($\delta >0.15)$, there is no evidence for any changes in
the low-energy gap structure at $T_{c}$ from either photoinduced QP
absorption or QP relaxation time data. In combination with the sum rule,
this implies the existence of pre-formed pairs up to a much higher
temperature $T^{*}.$ Near $\delta =0.1$ a rapid cross-over is observed, to a
state where the QP\ recombination time diverges and the photoinduced QP
density falls to zero at $T_{c}$, indicating the existence of a
temperature-dependent superconducting gap which closes at $T_{c}$.
\end{abstract}

Combining the universal correlations between $T_{c}$, the ratio of carrier
density to effective mass $n_{s}/m^{\ast }$ and pseudogap behaviour in the
underdoped region Uemura \cite{Uemura} suggested that the phase diagram of
high-temperature superconducting cuprates could be described in terms of a
crossover from Bose-Einstein (BE) to BCS-like condensation. According to
this\ scenario, in the underdoped state hole pairs form at temperatures
above $T_{c}$ which subsequently form a phase-coherent superconducting
condensate at $T_{c}$ with no associated change of pairing amplitude at this
temperature, while in the BCS-like state - suggested to exist in the
overdoped cuprates - an energy gap, phase coherence and superconductivity
would all occur more or less simultaneously close at $T_{c}$ as a result of
\ a collective phenomenon.

Using time-resolved (TR) optical spectroscopy to measure quasiparticle (QP)\
relaxation dynamics in YBa$_{2}$Cu$_{3}$O$_{7-\delta }$ (YBCO) on the
femtosecond timescale, we present experimental evidence for such a crossover
near optimum doping from a BE-like superconductor with pre-formed pairs in
the underdoped state to a BCS-like superconductor which shows a gap opening
close to $T_{c}$. The photoinduced optical transmission through a
superconductor thin film is proportional to the photoexcited QP density, so
it can be used to probe changes in the low-energy excitation spectrum.
Time-resolved measurements also provide independent direct information about
the gap from analysis of the QP recombination time, where for a
superconductor $\tau _{s}\propto 1/\Delta (T)$\cite{Rothwarf and Taylor}.
Consistent with appearance of a superconducting gap, previous TR
measurements on optimally doped cuprate superconductors have shown large
changes in both the photoinduced optical constants\cite{Han,Stevens} and
relaxation time\cite{Easley} near $T_{c}$. In this paper the data for a
large range of O doping $\delta $ are presented for YBa$_{2}$Cu$_{3}$O$%
_{7-\delta }$ and analysed using a theoretical model for the temperature
dependence of the photoinduced QP density\cite{Kabanov}, giving systematic
and quantitative information about the evolution of the energy gap with $%
\delta $ and $T$ in underdoped and near optimally doped samples.

The experiments were performed on a number of $\sim $100 nm thick epitaxial
YBCO films on either SrTiO$_{3}$ or MgO substrates which were annealed to
obtain $\delta $ in the range $0.48<\delta <0.1$. AC magnetic susceptibility
measurements show narrow (single)\ superconducting transitions of less than
2 K for $\delta \sim 0.1$ and 4-7 K for $\delta >0.2$ (defined by a 90\%
drop in the real part of the susceptibiliy $\chi ^{\prime })$. The optical
processes involved in the photoinduced experiments are shown schematically
in {\it fig}.1. First carriers are excited by the absorption of a 200 fs
pump laser pulse (step 1 in {\it fig}.1). After photoexcitation, the
carriers relax to QP\ states near the Fermi level within $10\sim 100$ fs by
carrier-carrier scattering and carrier-phonon scattering (step 2)\cite
{Allen,Brorson,Chekalin}. Subsequently, QPs accumulate above the energy gap,
forming a near-equilibrium distribution of QPs due to a relaxation
bottleneck. This bottleneck occurs because the gap in the low-energy
spectrum limits further QP\ energy relaxation to recombination processes
which include only high--energy phonons with energy greater than the gap\cite
{Rothwarf and Taylor,Kabanov}. This near-steady state QP population, which
is intimately related to the properties of the gap itself, can be detected
by a second, suitably delayed laser pulse (step 3). To maximize the induced
absorption signal, we use 200 fs pulses with a laser wavelength of 800 nm
(1.5 eV photon energy) from a Ti:Sapphire laser, which coincides with a O-Cu
charge transfer optical resonance in YBCO\cite{Stevens,PC}. The initial
states for the probe pulse absorption are QP states at $E_{F}$ and the final
unoccupied states are in a band 1.5 eV above $E_{F}.$ The probe pulse was
polarized in the $a-b$ plane and hence probes primarily the CuO-plane
excitations. The number of\ QPs photoexcited by each laser pulse is
approximately $n_{Q}=n_{p}\frac{E_{p}}{2\Delta (0)},$ where $\Delta (0)$ is
the gap magnitude and $n_{p}$ is the number of absorbed photons in each
laser pulse. With $\Delta (0)=$ 30 meV and with a pump power of typically 10
mW, we calculate the photoinduced carrier density to be $n_{Q}=2\times
10^{19}\sim 3\times 10^{20}$ cm$^{-3}$. Since the superconducting carrier
density is $n_{s}\sim 5\times 10^{21}$cm$^{-3},$ $n_{Q}$ $\ll n_{s}$ the
photoexcited QPs are only a very small perturbation on the system.

A typical normalized photoinduced transmission signal $\delta {\cal T}/{\cal %
T}\,$ as a function of time delay below and above $T_{c}$ on a
near-optimally doped sample with $\delta =0.1$ is shown in {\it fig}.1. The
risetime - which is characteristic of the establishment of the QP\
near-equilibrium state - is not resolved with our 200 fs pulses, while the
decay ranges from 300 - 3000 fs and characterises the QP recombination time.
A longer-lived decay component, which was already reported in optimally
doped samples\cite{Stevens} is also observed at all doping levels, but will
not will be discussed further in this paper. The normalized peak amplitude
of the observed photoinduced transmission amplitude $\left| \delta {\cal T}/%
{\cal T}\right| \,$ as a function of $T$ for different $\delta $ is shown in 
{\it fig}. 2a). Near optimum doping ($\delta \sim 0.1)$, the signal
amplitude drops rapidly and nearly dissappears close to $T_{c}$. With $%
\delta $ $>0.1$ however, in spite of the fact that $T_{c}$ is lower, the
amplitude drops at progressively {\em higher} temperatures above $T_{c}$. To
fit the temperature dependence of the induced absorption signal, we use the
theoretical formula derived for the photoinduced QP\ absorption \cite
{Kabanov}. For underdoped samples ($\delta >0.1)$ we assume the gap is
temperature independent, and then the induced transmission is given by:

\[
-\delta {\cal T}/{\cal T\propto }\frac{E}{\Delta }\left[ 1+\frac{2\nu }{%
N(0)\Omega _{D}}\exp (-\Delta /k_{B}T)\right] ^{-1} 
\]
where ${\cal E}$ is the pump excitation energy density per unit cell, $\nu
\sim 8$ is the number of phonon modes participating in the relaxation, $%
N(0)=2.2$ eV$^{-1}$spin$^{-1}$cell$^{-1}$ and $\Omega _{D}=0.1$ eV is a
typical phonon frequency. The value of $\Delta $ used in the fits shown by
the solid lines in Figure 2a) is given by $2\Delta =\beta k_{B}T^{*},$ where 
$T^{*}$ is defined as the temperature at which the amplitude $\left| \delta 
{\cal T}/{\cal T}\right| $ drops to 5 \% of its maximum low-temperature
value and $\beta =5\pm 1$. Although the choice of $\nu $ is somewhat
arbitrary, the fit is relatively insensitive to its actual value and only
has a small effect on the magnitude of $\beta $. The data for the underdoped
samples are seen to be in good agreement with the predicted temperature
dependence of $\left| \delta {\cal T}/{\cal T}\right| \,\,$ with a constant $%
\Delta $. However, attempting to use the formula above to describe the
near-optimally doped sample $\delta =0.1$ proves impossible as shown by the
dashed line in {\it fig}.2a). If, however we assume instead that $\Delta $
is tempereture {\em dependent,}\ \ and using a gap $\Delta =\Delta _{BCS}(T)$
of BCS\ form with 2$\Delta _{BCS}(0)=10k_{B}T_{c}$ (not $T^{*}$), the
agreement\ with the data becomes again very good over a wide range of
temperatures below $T_{c}$ (solid line in {\it fig}.2a)). We note that the
slight cusp in the data below $T_{c}$, which is present in all optimally
doped samples can only be reproduced by using a temperature-dependent gap $%
\Delta (T)$ which closes at $T_{c}$. Above $T_{c}$, where $\Delta
_{BCS}(T)=0,$ a small $\left| \delta {\cal T}/{\cal T}\right| $ signal
remains, which presumably corresponds to some remains of a $T$-independent
gap. We conclude that the underdoped and near-optimally doped sample data
cannot be described by the same form of temperature dependent gap, but a
cross-over from a $T$-independent gap to a dominant $T$-dependent gap occurs
near optimum doping. The $T^{*}$ obtained from {\it fig}.2a) are plotted
together with $T_{c}$ as a function of $\delta $ on the phase diagram in 
{\it fig}. 2b). They coincide rather well with the ''pseudogap'' temperature 
$T_{p}$ where a drop in $N(0)$ has been deduced from NMR \cite{NMR},
infrared \cite{IR} and specific heat\cite{Loram} measurements amongst others.

Turning our attention to the QP recombination time $\tau _{s}$ obtained from
a single exponential fit of the data near optimum doping, a sharp divergence
is observed just below $T_{c}$ which then drops to a near-constant value at
low temperatures ({\it fig}.3 a)). From the existence of this divergence in $%
\tau _{s}$ and the fact that below $T_{c},$ $\tau _{s}\propto 1/\Delta (T)$%
\cite{Rothwarf and Taylor,Kabanov} we can unambiguously deduce the presence
of a gap $\Delta (T)$ which {\em closes at }$T_{c}$. With increasing $\delta 
$ however, this divergence of $\tau _{s}$\ near $T_{c}$ rapidly dissappears.
As shown in {\it fig.}3b) for $\delta =0.3$, $\tau $ is nearly constant with
temperature above and below $T_{c}$ and shows no visible anomaly at $T_{c}$
or $T^{*}$({\it fig}.3b)). A plot of the normal state relaxation time $\tau
_{n}$ for $T>T_{c}$ (at 100 K) and $\tau _{s}$ for $T<T_{c}$ (at 20 K) as a
function of $\delta $ in {\it fig}. 3 c) shows a step-like change of $\tau
_{s}$ near optimum doping clearly suggesting the existence of a cross-over
in behaviour.

Let us now discuss the possible origins of the observed cross-over. Near
optimum doping the gap appears to be formed by a {\em collective effect}
implying a BCS-like scenario where the creation of pairs occurs
simultaneously with macroscopic phase coherence and the opening of a gap.
With such a BCS-like gap $\Delta _{BCS}(T)$ we can explain quantitatively
the divergence of $\tau _{s}$ at $T_{c}$ (Fig.3a)) and the temperature
dependence of $\left| \delta {\cal T}/{\cal T}\right| \,$({\it eq}.(1) and 
{\it fig}. 2a)) together with the fact that $\delta {\cal T}/{\cal T}%
\rightarrow 0$ at $T_{c}$ near optimum doping.

In the underdoped state with $\delta >0.15$, the anomaly of $\tau _{s}$ at $%
T_{c}$ is no longer visible. Noting that energy conservation together with
the sum rule dictate that pairing {\em must }involve a change in the single
particle DOS$,$ the absence of anomaly in $\tau $ at $T_{c}$ together with
the relation $\tau \propto 1/\Delta $ imply that there is no change in the
pairing amplitude at $T_{c}$ itself. Moreover, within experimental error, $%
\left| \delta {\cal T}/{\cal T}\right| \,$also shows no change at $T_{c}$,
which, considering {\it eq}. (1) also implies that there is no change in the
DOS\ at this temperature$^{(1)}$ and certainly that no gap opens {\em at }$%
T_{c}$. In fact, from the $T$-dependence of both $\left| \delta {\cal T}/%
{\cal T}\right| \,$and $\tau _{s}$\ ({\it figs}. 2a) and 3b) respectively),
the gap appears to be more or less $T$-independent.

Let us now consider these underdoped sample data in the gereneralised
framework of Boson condensation superconductivity theories, where the
experimentally verifyable consequences arise from the fact that no
collective pairing effect occurs at $T_{c},$ but rather bosons, which are
already present at $T_{c}$ condense into a phase-coherent macroscopic
superconducting ground state without any change in pair density at this
temperature. In the BEC scenario as applied to the underdoped cuprates,
pairing starts to take place well above $T_{c}$ and is governed by the
thermal excitations from the pair ground state at {\sf E}$_{0}$ to single
particle states {\sf E}$_{1}$ at an energy $2\Delta $ above it ({\it fig}%
.1). The gap 2$\Delta $ now signifies the local pair binding energy $E_{B}$
and is $T$-independent, consistent with the $T$-dependence of $\left| \delta 
{\cal T}/{\cal T}\right| \,$and $\tau $. Its magnitude\ should decrease with
increasing doping because of screening\cite{Alexandrov}, which is also
consistent with the data on $T^{*}$ in {\it fig}. 2b). This scenario with a
constant splitting between the pair ground state and the unpaired excited
state thus explains why $\left| \delta {\cal T}/{\cal T}\right| \,$does not
vanish at $T_{c}$ in underdoped samples, but is dominated by ''pseudogap''
behaviour with asymptotic temperature dependence above $T_{c}$ resulting
from the statistics of the 2-level system\cite{Kabanov}.

To understand better the behaviour of photoexcited charge carriers in this
scenario, let us consider the relaxation process in more microscopic terms.
After inital photoexcited {\it e-h} pair relaxation by {\it e-e} scattering
and phonon emission - a process which is insensitive to the low-energy
structure - the particles end up in {\it single particle states} near $E_{F}$%
. The next relaxation step involves pairing of these carriers into the pair
ground state with the release of $E_{B}$ per pair. This pairing process is
similar as for QP recombination in the BCS case, except that the gap $\Delta
_{BCS}(T)$ is now replaced by a $T$-independent $E_{B}$\cite{Kabanov}.
Because of the large dielectric constant in YBCO at low frequencies ($%
\varepsilon _{r}>$100\cite{Timusk}) the Coulomb repulsion for carriers on
adjacent sites is small compared to either the typical bipolaronic binding
energy or the exchange interaction energy $J$. Thus pairs on adjacent
lattice sites can form relatively easily, irrespective of whether the
pairing mechanism is bipolaronic, electronic, or some combination of the
two. Since the in-plane coherence length $\xi >a,$ where $a$ is the
intersite carrier spacing, to first approximation the pairs can be treated
as local bosons, which is a condition for BEC\ to occur. Macroscopic phase
coherence and condensation in the superconducting ground state occurs when
thermal phase fluctuations of such bosonic pairs are sufficiently reduced to
enable phase locking between neighbouring boson wavefunctions. This phase
locking at $T_{c}$ is manifested by the appearance of a macroscopically
coherent Meissner state and zero resistivity, but is not visible in the
pairing amplitude or in the single-particle DOS, and hence also not
measurable in $\delta {\cal T}/{\cal T}$ or $\tau _{s}$. By heuristic
argument, phase coherence should occur when the pair deBroglie wavelength $%
\lambda $ becomes comparable to the coherence length $\xi .$ Estimating $%
T_{c}$ from $k_{B}T_{c}\sim h/(m^{\ast }\xi ^{2}),$ with $m^{\ast }=3m_{e}$
and $\xi =18$\AA\ (in-plane for YBCO) we obtain $T_{c}\sim $100K. More
rigorously, applying the Bose-Einstein formula $T_{c}\simeq \hbar
^{2}n_{s}^{2/3}/(2m^{\ast })k_{B}$ to YBCO with the same $m^{\ast }=3m_{e}$
and $n_{s}\sim 10^{21}$cm$^{-3}$ gives values consistent with the observed $%
T_{c}$-s\cite{Uemura}.

We conclude that the distinct differences of the low-energy carrier dynamics
in YBCO underdoped state compared to the optimally doped state imply the
existence of a cross-over near $\delta \sim 0.1$ from a state in which
pre-formed bosonic pairs condense at $T_{c}$ with no observable change in
the DOS at this temperature, to a BCS-like state exhibiting clear anomalies
related to a gap opening at $T_{c}$. Throughout the discussion of the
present data we have discussed the concept of condensation in general terms
and intentionally kept it separate from the issue of the microscopic pairing
mechanism. For a recent survey of some of the high-$T_{c}$ theories which
rely on condensation for achieving a macroscopically coherent
superconducting ground state - ranging from the generalized Hubbard model to
bipolaronic superconductivity - we refer the reader to ref.\cite{Micnas}.

We wish to thank V.V.Kabanov for his comments and critical reading of the
manuscript.

\section{Footnote}

(1) A similar conspicuous absence of anomalies at $T_{c}$ in underdoped
samples is also observed in other experiments which probe the DOS, for
example infrared spectroscopy\cite{IR,Alexandrov2}, specific heat\cite
{Loram,Alexandrov2} and single-particle tunneling\cite{Deutscher} etc.

\section{Figure Captions}

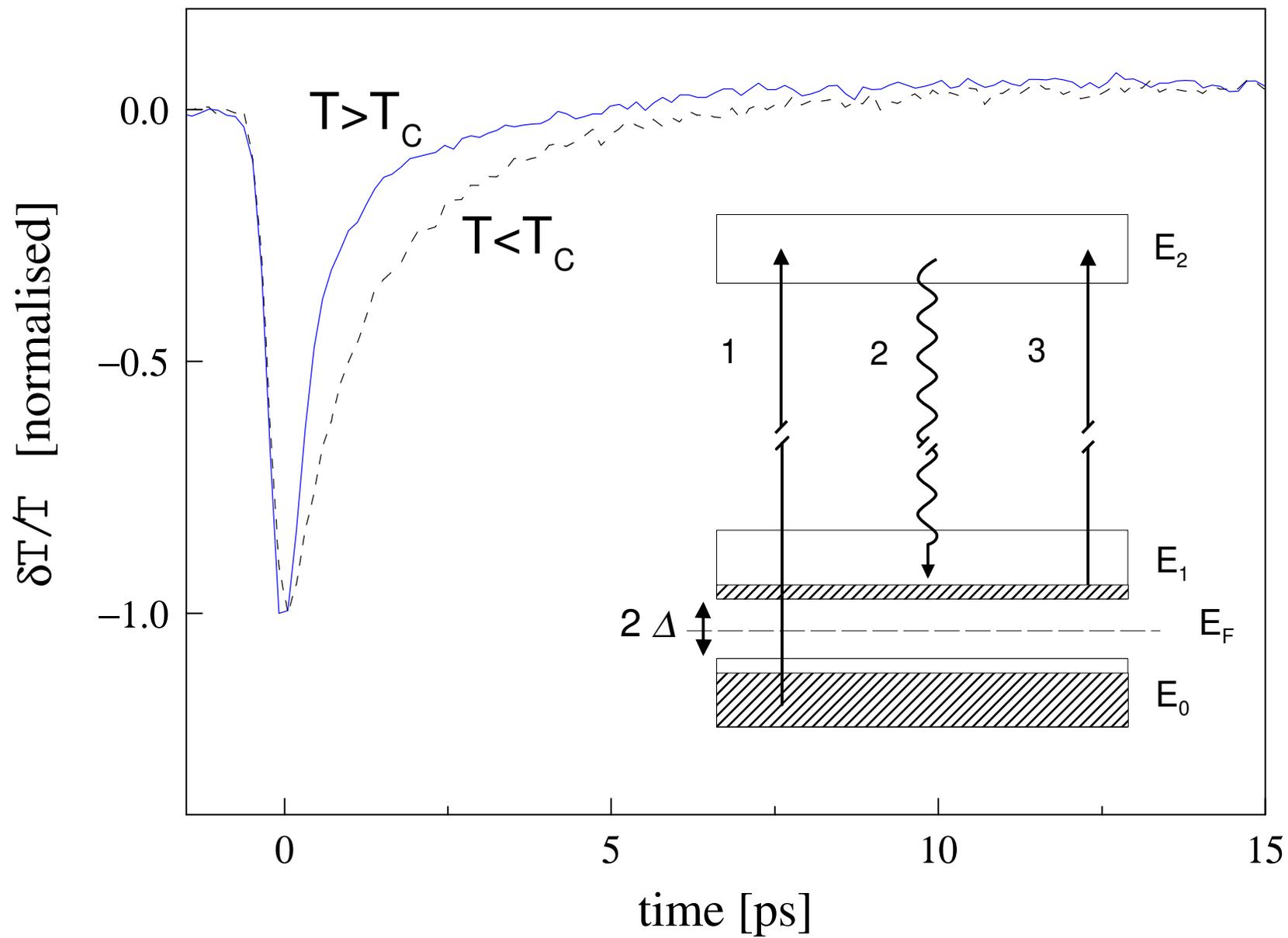
Figure 1. The normalised time-resolved induced transmission as a function of
time delay just above and just below $T_{c}$\ for YBCO with $\delta =0.1$.
The schematic diagram shows the optical absorption and QP relaxation
processes involved in the experiment.

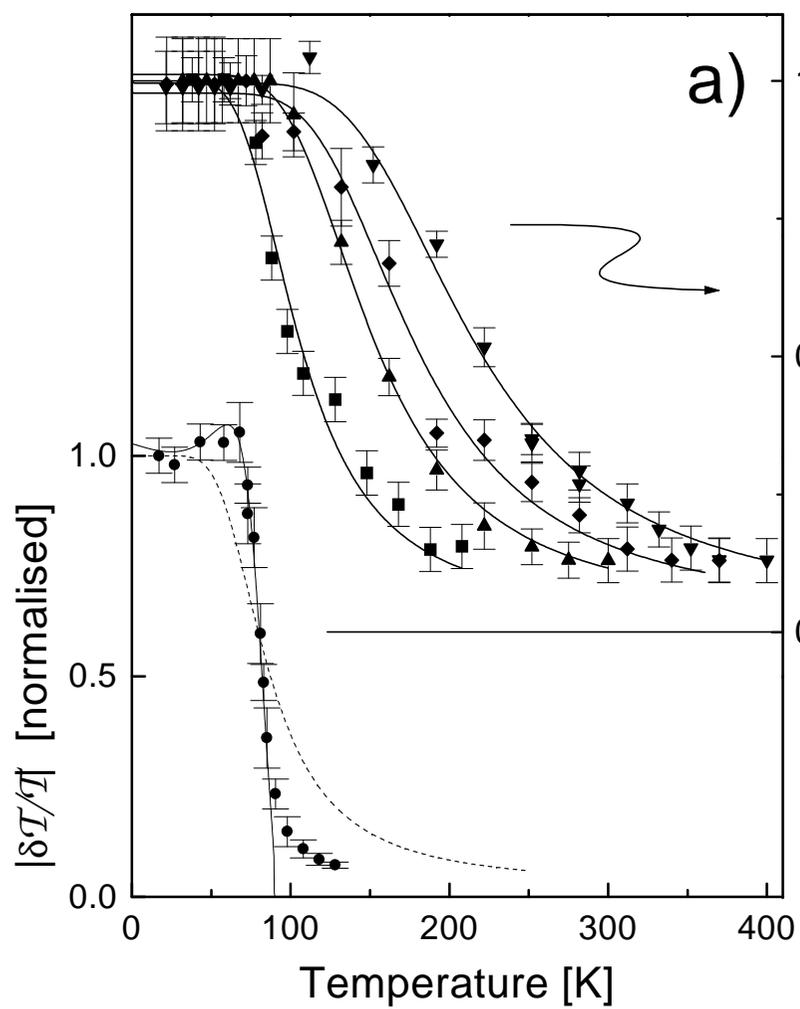
Figure 2. a) The normalized amplitude of the induced transmission $\Delta 
{\cal T}$/${\cal T}$ \ for near-optimally doped sample with $\delta \sim
0.05,T_{c}=90$K (circles) and for underdoped samples with $\delta =0.14$, $%
T_{c}=84$K (squares), $\delta =0.18$, $T_{c}=77$K (up triangles), $\delta
=0.44,T_{c}=53$K (diamonds) and $\delta =48$, $T_{c}=48$K (down triangles)
respectively. The underdoped sample data are displaced for clarity. The
lines are theoretical fits to the data from the model of Kabanov et al \cite
{Kabanov}. b)\ $T^{\ast }$ (full circles) and $T_{c}$ (line and squares) as
a function of $\delta $.

Figure 3. $\tau _{s}$ as a function of $T$ for a)\ $\delta =0.05$ ($T_{c}$ =
90K) and b) $\delta =0.3$ ($T_{c}$ = 60K). The dashed line in Fig.a) is a
fit with $\tau =A/\Delta (T)$, where $A$ is a constant and $\Delta (T)$ is a
BCS-like $T$-dependent gap. c)\ The relaxation time $\tau $ as a function of 
$\delta $ for $T>T_{c}$ and $T<T_{c}$ $($at $T=100$ K$\,$and 20 K
respectively$).$ The relaxation time $\tau _{s}$ in the superconducting
state shows a cross-over near optimum doping, while $\tau _{n}$ ($T>T_{c}$)
is nearly doping and temperature independent.

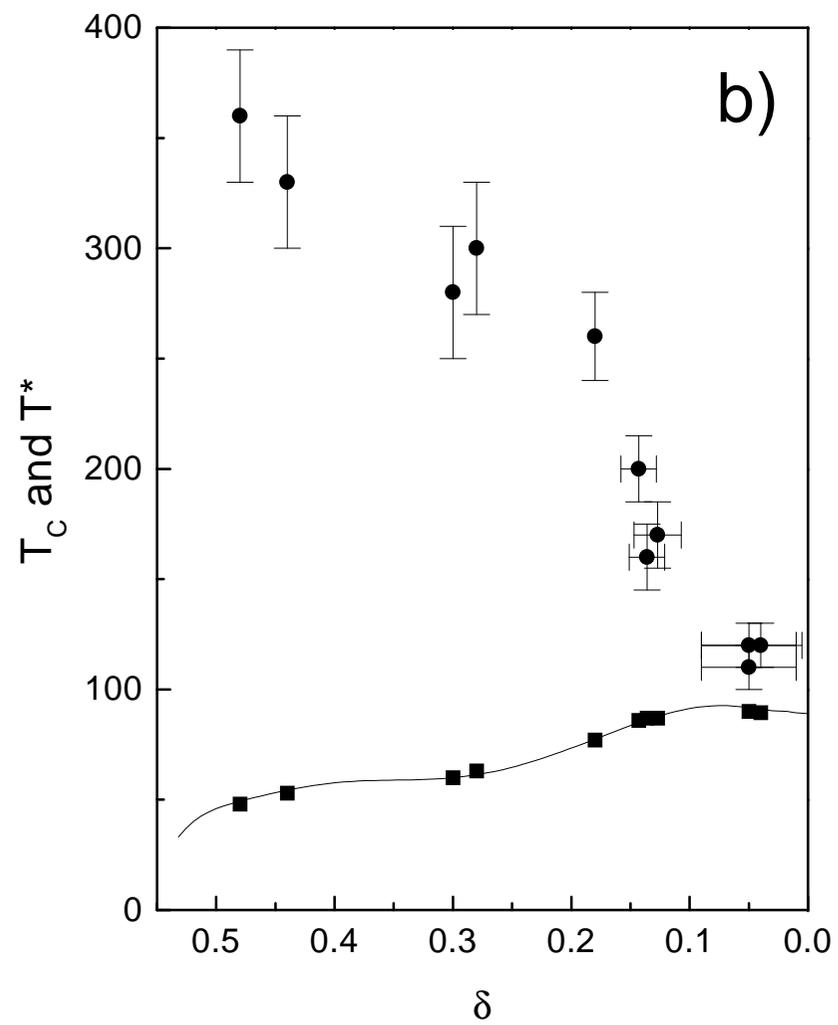
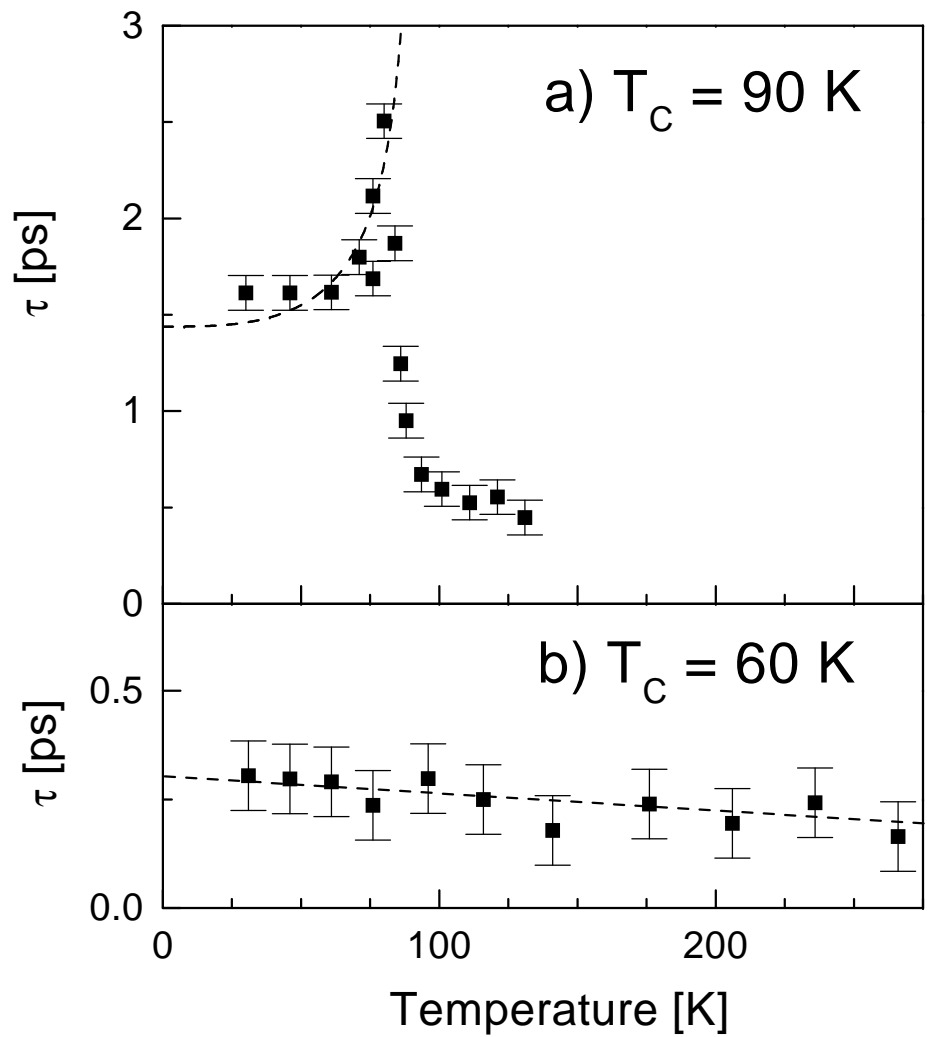
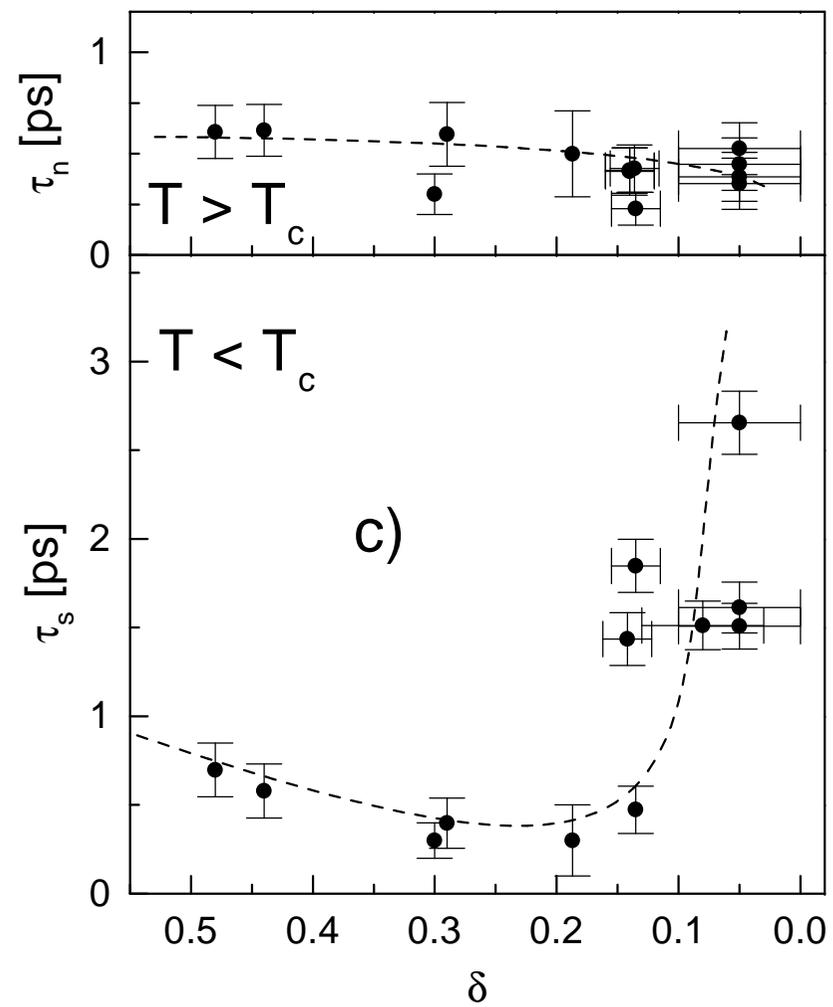
\end{document}